\documentclass[12pt,preprint]{aastex}

\usepackage{color}
\usepackage{amsmath}
\usepackage{float}
\usepackage{natbib}

\def\mj{$\,${\rm M}$_{\rm J}\,$}
\def\me{$\,{\rm M}_{\oplus}\,$}
\def\h2o{H$_2$O}
\def\sio2{SiO$_2$}
\def\gc3{g\,cm$^{-3}$}
\def\swr{Schwarzschild }
\def\ldx{Ledoux }

\begin{document}

\title{Convection and Mixing in Giant Planet Evolution}

\author{A.~Vazan, R.~Helled, A.~Kovetz, M.~Podolak}
\affil{Department of Geophysics, Atmospheric, and Planetary Sciences 
   Tel-Aviv University, Israel}



\begin{abstract}
The primordial internal structures of gas giant planets are unknown. Often giant planets are modeled under the assumption that they are adiabatic, convective, and homogeneously mixed, but this is not necessarily correct. In this work, we present the first self-consistent calculation of convective transport of both heat and material as the planets evolve. We examine how planetary evolution depends on the initial composition and its distribution, whether the internal structure changes with time, and if so, how it affects the evolution. We consider various primordial distributions, different compositions, and different mixing efficiencies and follow the distribution of heavy elements in a  Jupiter-mass planet as it evolves. 
We show that a heavy-element core cannot be eroded by convection if there is a sharp compositional change at the core-envelope boundary. If the heavy elements are initially distributed within the planet according to some compositional gradient, mixing occurs in the outer regions resulting in a compositionally homogeneous outer envelope. Mixing of heavy materials that are injected in a convective gaseous envelope are found to mix efficiently. Our work demonstrates that the primordial internal structure of a giant planet plays a substantial role in determining its long-term evolution and that giant planets can have non-adiabatic interiors. 
These results emphasize the importance of coupling formation, evolution, and internal structure models of giant planets self-consistently.

\end{abstract}


\section{Introduction}
The existence of a core in giant planets and its physical properties has important consequences for our understanding of planet formation and internal structure \citep{Helled2014}. 
Most investigations of giant planet structure and evolution have assumed that the planet consists of several distinct regions which remain chemically homogeneous during the long-term evolution \citep{baraffe08, fortney07, vazan13}. This is, however, a simplifying assumption and in reality it is possible that different regions of the interior will mix or separate as the planet evolves.  Changes in composition can also arise when one material becomes soluble in another. Recent studies of the solubility of analogous phases in the planetary interior have shown that \h2o \citep{wilson12b}, \sio2 \citep{wilson12a}, and iron \citep{wahl13} dissolve in liquid metallic hydrogen at high temperature and pressure. These calculations imply that giant planet interiors are in thermodynamic disequilibrium with surrounding layers, promoting redistribution of heavy elements. 

Alternatively, as the planet evolves parts of the interior may enter the pressure-temperature regime where one material is immiscible in another.  An example of this would be helium rainout during the evolution of a gas giant \citep{stevsal77, forthub03}.  This leads to a helium-rich region just outside the core.   
On the other hand, \cite{stevenson82a} has suggested that (part of) the planet's core can be soluble in the surrounding gaseous envelope, and that during the formation processes, some of this core material might be dredged up and mixed with the surrounding hydrogen and helium.
This effect, however, is expected to be small \citep{lisstev07}. 

The distribution of the heavy elements in the planet, whether concentrated in a central core or mixed into a hydrogen-helium envelope, can be of great importance for determining planetary thermal evolution and has consequences for the total amount of heavy elements in the planet deduced from static models  \citep{stevenson85,chabr+barf07,lecontechab12,lecontechab13}.  Rearrangement of the heavy elements has important consequences for the understanding of the planet formation process. For example, core erosion and mixing could explain how massive planets formed by core accretion can have small cores. Such a process might also explain, for example, why Jupiter appears to have a smaller core than Saturn \citep{stevenson82a, guill04}. Furthermore, composition redistribution also changes the moment of inertia of the planet, making it harder to deduce the formation mechanism even with the more precise measurements expected in the future \citep{helled12}.

An additional assumption that is often made is that there is efficient large-scale convection in giant planets. However, there are several situations where this assumption is incorrect \citep{guill94, saum+guil04, fortney11}. The existence of a core or other gradient in the distribution of heavy elements, in particular, weakens the adiabatic assumption. When a stabilizing composition gradient exists, there is no large-scale convection, and the efficiency of the heat transport can strongly be reduced. Thus, while thermal evolution models have been calculated for giant planets, these models generally consider the effect of convection on the evolution of the temperature gradient, but not on the evolution of the heavy element distribution \citep{fortney11, chabr+barf07, mirouh12}. \cite{lecontechab12, lecontechab13} emphasize the importance of a composition gradient on the thermal evolution of the planet, but do not actually follow the evolution of the composition gradient. Since convection transfers both heat and material, and since the convection itself depends on both the thermal and compositional gradient, these two transport processes must be considered simultaneously. 

In this work we present the first self-consistent calculation of convective transport of both heat and material.  The criterion for the onset of convection is taken to be the \ldx criterion. The \swr criterion is also considered for comparison.  In each case, within the convective regions the transport of both heat and material is computed using the mixing length recipe.  We then calculate full evolutionary models and demonstrate how the compositional structure changes with time, and how this affects the radius and luminosity of giant planets as they evolve. 

\section{The method of calculation}

The structure and evolution calculations are carried out using a planetary evolution code that solves the equations

\begin{eqnarray}
{\partial\over\partial m}{4\pi\over3} r^3&=&{1\over\rho}, \label{r_eq} \\
{\partial p\over\partial m}&=&-{Gm\over4\pi r^4}, \label{h.eq} \\
{\partial u\over\partial t}+p{\partial\over\partial t}{1\over\rho}&=&-{\partial L\over\partial m}, \label{energ_eq} \\
{\partial T\over\partial m}&=&\nabla{\partial p\over\partial m}, \label{T_eq} \\
{\partial X_j\over\partial t}&=&-{\partial F_j\over\partial m}, \label{X_eq} 
\end{eqnarray}
where $\nabla=d\ln T/d\ln p$ is the 'temperature gradient', which may be radiative (and conductive) or convective, depending on the planet's local properties (Appendix~A).
In the last equation, $X_j$ is the mass fraction of the $j$'th species, and 
\begin{equation}
F_j=-(4\pi r^2\rho)^2D{\partial X_j\over\partial m}
\label{X_flux} \end{equation}
is the corresponding particle flux through the sphere containing mass $m$. The convective diffusion coefficient $D$ (Appendix~A) is non-zero in convective regions, otherwise it vanishes. 
The boundary conditions at the center are
\begin{equation}
m=r=L=F_j=0 .
\label{cbc} \end{equation}
At the surface, which we choose as the photosphere,
\begin{equation}
m=M,\quad \kappa p_G=(GM/R^2)\tau_s,\quad L=4\pi R^2\sigma T^4,\quad F_j=0, 
\label{sbc} \end{equation}
where $\tau_s$, the optical depth of the photosphere, is set to be 1.

Eqs.~(\ref{r_eq})--(\ref{X_eq}) are replaced by difference equations on a grid of $n\sim100$ points. In order to get an optimal grid resolution during evolution, we add the second-difference equation
\begin{equation}
f_{i+1}-f_i = f_i-f_{i-1},\qquad i=2,...,n-1\label{dbldiff}\end{equation}
where
\begin{equation}
f = (m/M)^{2/3}-c_1\ln p+c_2\ln(r^2+c_3),
\label{subdiv} \end{equation}
and the $c$'s are suitably chosen, positive, constants. Eq.~(\ref{dbldiff}) forces equal steps of $m^{2/3}$ near the center, and equal steps of $\ln p$ or $\ln r$---whichever step is larger---in the outer region.
The difference equations
are {\it all} solved simultaneously for each time step. Further details of the code can be found in \cite{kovetz09}.

In this work we use the equations of state tables from the work of \cite{scvh} for hydrogen and helium. For the high-Z material, which is assumed to be either ``ice'' (\h2o) or ``rock'' (\sio2), we computed an equation of state based on the quotidian equation of state (QEOS) of \cite{qeos88}, as described in \cite{vazan13}.
The opacity is computed using the radiative opacity tables of \cite{pollack85} for solar-composition, and the thermal conductivity tables of \cite{potekhin99}.

Unless otherwise stated, the onset of convection in the planetary interior during evolution is determined by the Ledoux's criterion, 
\begin{equation}
\Delta \nabla \equiv \nabla_R - \nabla_A - \nabla_X >0,
\label{eq_led} \end{equation}
where $\nabla_R$ and $\nabla_A$  are the radiative and adiabatic temperature gradients, respectively; and 

\begin{equation}\label{nabledoux}
\nabla_X=\sum_j \frac{\partial \ln T(\rho,p,X)}{\partial X_j}\frac{dX_j}{d\ln p}.
\end{equation}
If $\nabla_X=0$, convection sets in when $\nabla_R>\nabla_A$, the usual \swr criterion \citep{schwarz06}. 
Further details, in particular the calculation of $\nabla$ in eq.~(\ref{T_eq}), as well as that of $D$ in eq.~(\ref{X_flux}), are given in Appendix~A. 
We note that, in accordance with eq.~(\ref{eq_led}), steep enough composition gradients---in a direction that renders $\nabla_X$ positive---will cause convection to cease completely.
If, on the contrary, $\Delta \nabla \le 0$, the actual gradient is the radiative gradient $\nabla_R$ and the diffusion coefficient is zero: there is neither convective heat transport, nor any mixing. 
We do not consider semi-convection---that is, the situation $\nabla_A < \nabla_R < \nabla_A+\nabla_X$---or double-layered convection in this study. 
Thus, our solution (when using \ldx criterion) can provide a lower bound for the mixing efficiency inside the planet.

Although the mixing is mainly controlled by the initial composition gradients and the convection criterion, the rate of mixing can vary with the model parameters. 
One of the most important is the mixing length parameter, $\alpha=\ell/H_p$, which determines the rate at which the material is mixed, since it affects the convective diffusion
coefficient $D$. 
In this work we consider different values for $\alpha$, allowing it to change by several orders of magnitude. 
The planetary evolution and the mixing of heavy elements for different primordial internal structure and mixing length parameters are described below. 

\section{Heavy-element redistribution}

As we have noted above, the presence of convection and its strength depend on the temperature and composition distributions. 
In order to understand the consequences for different primordial internal structures, we consider three representative cases: (1) A planet with a core consisting of pure high-Z material, 
surrounded by an envelope of hydrogen and helium in solar ratio. (2) A planet where Z, the mass fraction of heavy elements, decreases continuously from a value of Z=1 at the center to Z=0 at the 
base of the envelope.
(3) A solar-mix hydrogen-helium planet where the high-Z material is distributed as a Gaussian around some point in the interior. 
This might be taken to represent the injection of material by a giant impact in the early history of the planet. Below we present the results for these cases. 
In all the calculations we consider a 1\mj planet. Unless otherwise noted the high-Z material is represented by water and the criterion for convection is Ledoux's.

\subsection{Core - envelope}
\begin{figure}[ht]
\centerline{\includegraphics[angle=0, width=14cm]{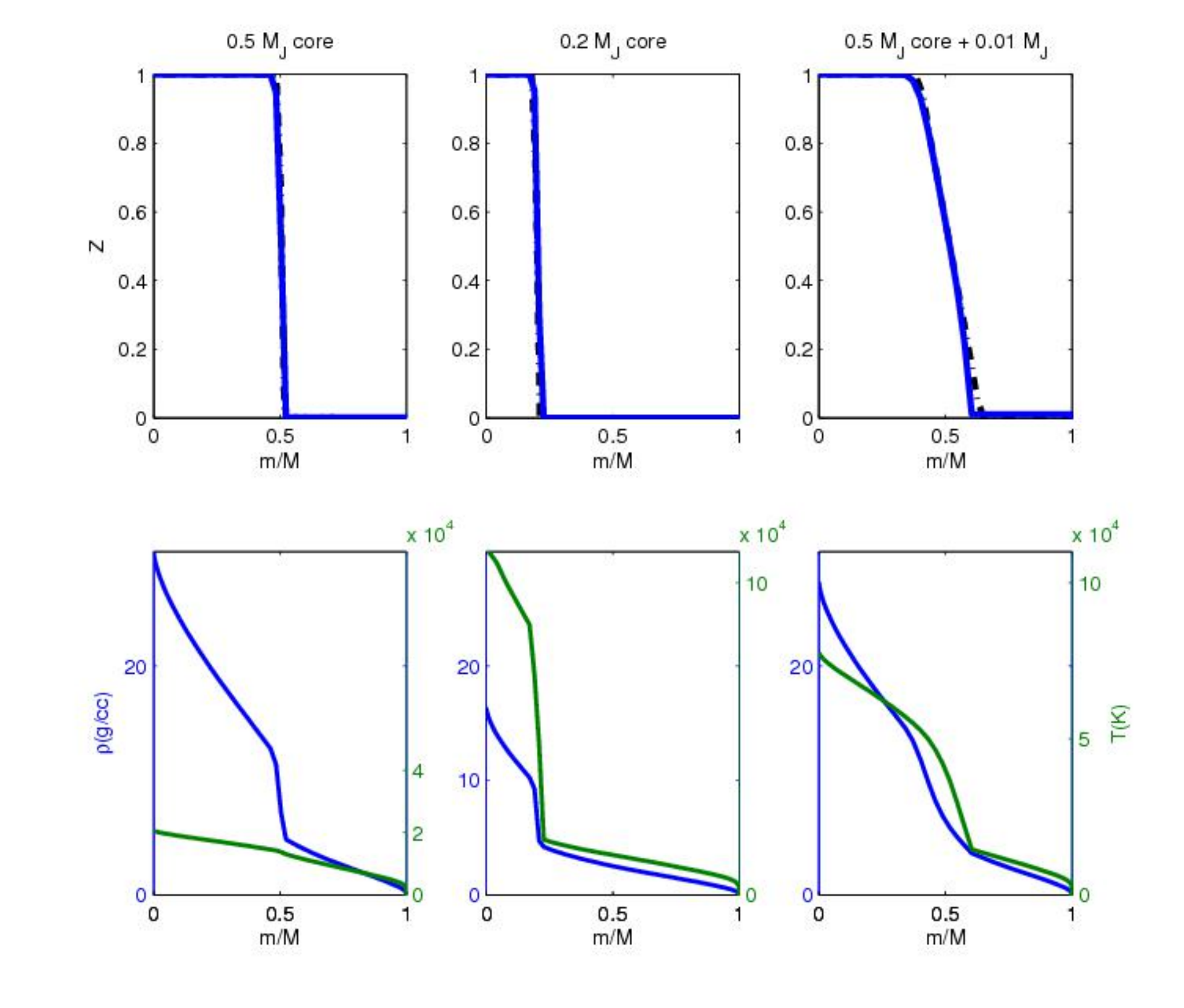}}
\caption{Heavy element (\h2o) mass fraction vs.~normalized planetary mass for a 1\mj planet after 10$^{10}$ years of evolution for a planet consisting of a high-Z core and a hydrogen-helium envelope. Initial internal structure (dashed dotted) and after 10$^{10}$ years of evolution (solid) are presented. Left: $M_c=0.5$\mj; Center: $M_c=0.2$\mj; Right: $M_c=0.5$\mj with a small Z-gradient above the core ($0.01$\mj). The bottom panels show the density (blue) and temperature (green) vs.~normalized mass for the different cases after 10$^{10}$ years. }\label{z_core}
\end{figure}

Fig.~\ref{z_core} shows results for three cases of a core-envelope structure. The black dash-dot curve represents the initial Z as a function of normalized mass inside the planet, 
while the blue curve shows the Z distribution after $10^{10}$ years of evolution.  Three different cases are presented: The first is a massive heavy element core of 0.5\mj, surrounded by a hydrogen-helium 
envelope (left).  The second is a smaller heavy element core of 0.2\mj, surrounded by a hydrogen-helium envelope (middle). The third is a massive heavy element core of 0.5\mj, surrounded by a hydrogen-helium 
envelope, but here we have mixed an additional 0.01\,\mj of high-Z material.  This material is distributed along a linear gradient starting at Z=1 near the core and going to Z=0 at $m/M=0.6$.  
This is meant to represent an initial state that results from some previous heavy-element settling process and/or diffusion. 
The bottom panels show the density (blue) and temperature (green) as a function of normalized mass for the different cases after $10^{10}$ years. The mixing length parameter in these 
simulations was $\alpha=0.5$, but the results are not sensitive to the exact value chosen since $\alpha$ only determines the rate at which the material mixes and not the actual onset of mixing.
   
In the case of a core-envelope configuration (left and middle panels), the sharp composition gradient stabilizes the planet against convection and a radiative layer is formed at the core-envelope boundary.  
As a result, although large parts of the envelope do convect, convection is inhibited near the core, and the distribution of heavy elements within the planet remains essentially unchanged. 
Even after $10^{10}$ years of evolution the outer envelope heavy element mass fraction is negligible. 
A consequence of this inhibition of convection is that the planet retains its heat longer. 
Therefore, the memory of the initial temperature profile, which is generally lost after $\sim 10^7$ years, can persist for as long as $\sim 10^{10}$ years.  
 
Lowering the mass of the core does not cause it to mix more readily.  This is demonstrated in the middle panel, where even a 0.2\mj core remains unmixed after $10^{10}$ years.  
However, second order processes, like diffusion or chemical solubility, might still slowly erode the core and establish a composition gradient in the envelope. 
In the right-hand panel of Fig.~\ref{z_core}, we simulate a case where a thin region with a composition gradient exists around the core. 
In this case, as shown in the figure, the outer part of this region becomes convective during evolution, and the high-Z material mixes uniformly through the outer envelope.  
This mixing, however, occurs only in the outermost part of the planet , where the thermal expansion coefficient is high, and does not affect the core directly.  
It therefore appears that core erosion cannot proceed via rapid upward mixing when there is a sharp compositional gradient, but only through much slower diffusive processes.

\subsection{Gradual composition gradient}  
\begin{figure}[ht]
\centerline{\includegraphics[angle=0, width=14cm]{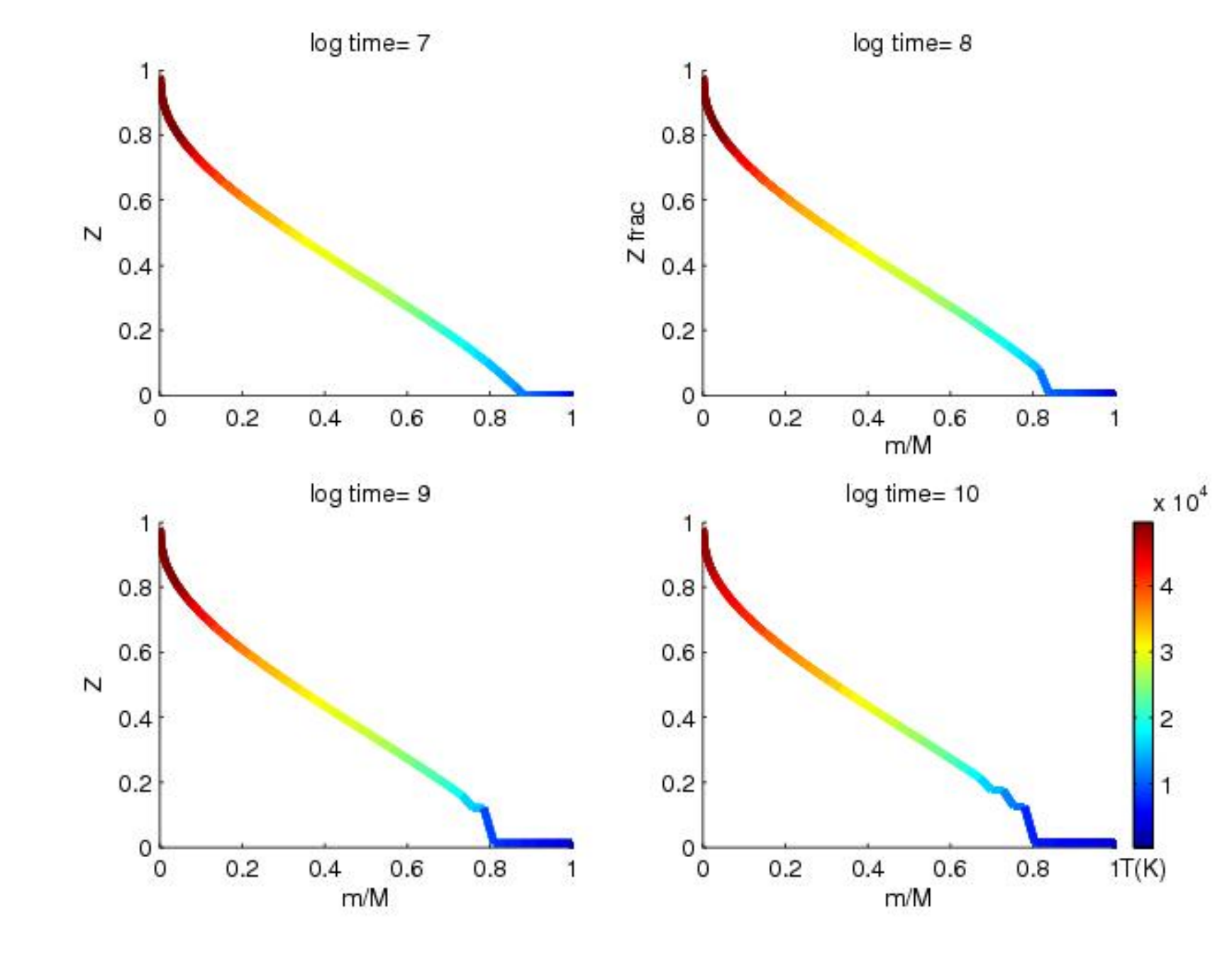}}
\caption{Evolution of heavy element (\h2o) distribution for 1\mj planet in time for a continuous Z-gradient at different times in the evolution.  The step structure (see text) can be seen developing.  The colors on each curve show the evolution of the temperature profile.}\label{z_grad}
\end{figure}

Once the composition gradient is reduced from being a sharp step to something more  gradual, convective mixing may come into play in different regions in the planet. 
A typical case 
is shown in fig.~\ref{z_grad}. The high-Z distribution is shown for different times in the evolutionary sequence. The color of the curves shows the associated temperature profile.

\begin{figure}[ht]
\centerline{\includegraphics[angle=0, width=14cm]{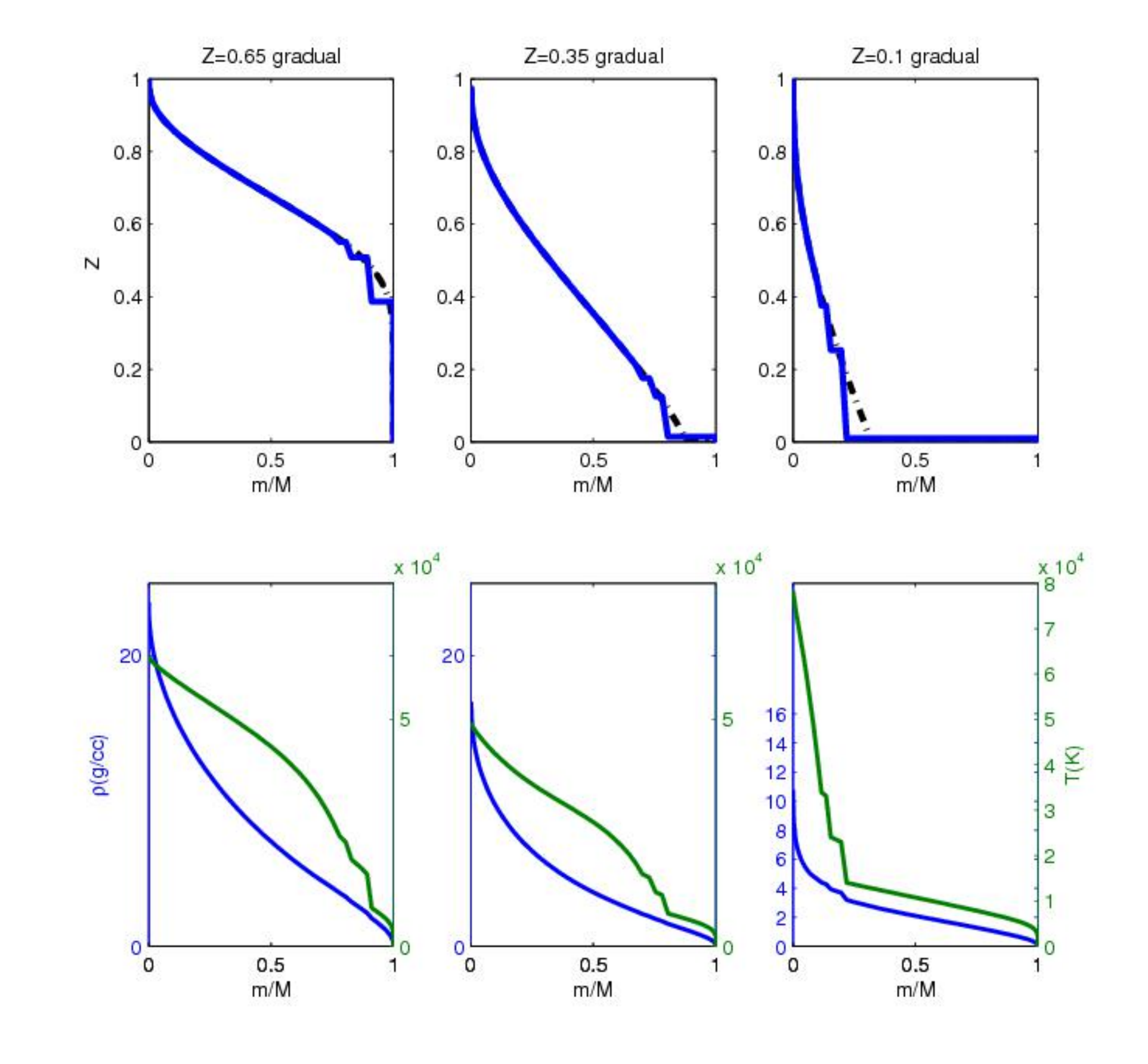}} 
\caption{Upper panels: Z as a function of $m/M$ after $~10^{10}$\,years of evolution of a 1\mj planet for different initial Z-gradients and total Z with the \ldx criterion. Shown are cases with Z=0.65 (left), Z=0.35 (center), and Z=0.10 (right). Lower panels: Density (blue) and temperature (green) after $\sim 10^{10}$\,years for the cases in the upper panels.}\label{compgrad}
\end{figure}

As can be seen from the figure, the upper 20\% of the mass is nearly completely mixed after $10^{9}$ years, while the innermost 80\% or so maintains its original Z-profile.  
There is an intermediate narrow region between the radiative and convective where convection is not completely suppressed. The steps in heavy element composition in this region  may be a  numerical artifact. 
In any case their effect on the subsequent structure and evolution is small,
as we show below (section 3.2.2).
In the outer regions convection becomes strong enough to completely mix the overlying material, and indeed the heat transport tracks the material transport closely.
The inner 80\% of the planet stays at roughly the same temperature while the outer 20\% cools as the mixing proceeds. 
Fig.~\ref{compgrad} shows the effect of changing the slope of the composition gradient.  
For all three panels Z=1 at the center and decreases linearly with radius to a surface value determined by the desired total Z for the planet. 
In all cases the original heavy element distribution is retained in the inner region of the planet while the outer part becomes homogeneously mixed.  

\subsubsection{Dependence on mixing length parameter}
\begin{figure}[ht]
\centerline{\includegraphics[angle=0, width=14cm]{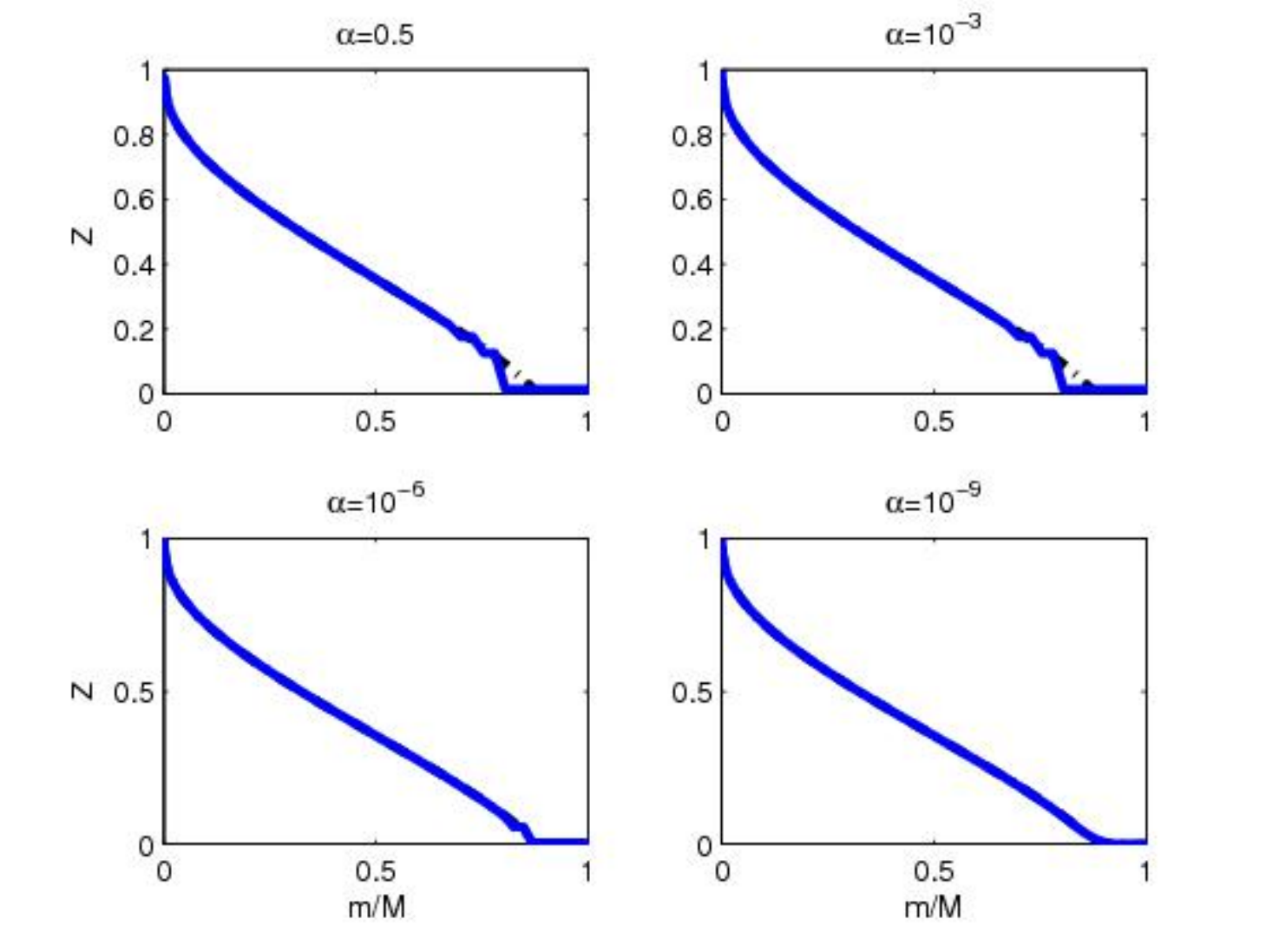}}
\caption{Z redistribution as a function of normalized mass after 10$^{10}$ years of evolution for the different values of mixing length parameter, $\alpha$.}\label{zalpha}
\end{figure}

The mixing length parameter used in Fig.~\ref{z_grad} and~\ref{compgrad} is $\alpha=0.5$. However, since the mixing rate depends on the value of $\alpha$, it is important to investigate how 
the mixing of the heavy elements in the planet changes for different values of $\alpha$. In stellar evolution models the mixing length is typically taken to be of the order of 
the pressure scale height, so that $\alpha\sim 1$.
In planetary interior conditions, however, the mixing parameter could be significantly smaller \citep[see, e.g., ][]{lecontechab12}.
Fig.~\ref{zalpha} shows how the choice of $\alpha$ affects the convective mixing during the planetary evolution.  For $\alpha=0.5$ (upper left) mixing is efficient and remains so for $\alpha\gtrsim 10^{-3}$.  
Lower values reduce the mixing efficiency until for $\alpha = 10^{-9}$ the internal compositional structure is essentially unchanged even after 10$^{10}$ years of evolution. 
For such low $\alpha$ values \citep{lecontechab12} radiative transfer is more effective than convection.

\subsubsection{Dependence on number of mesh points}
\begin{figure}[ht]
\centerline{\includegraphics[angle=0, width=10cm]{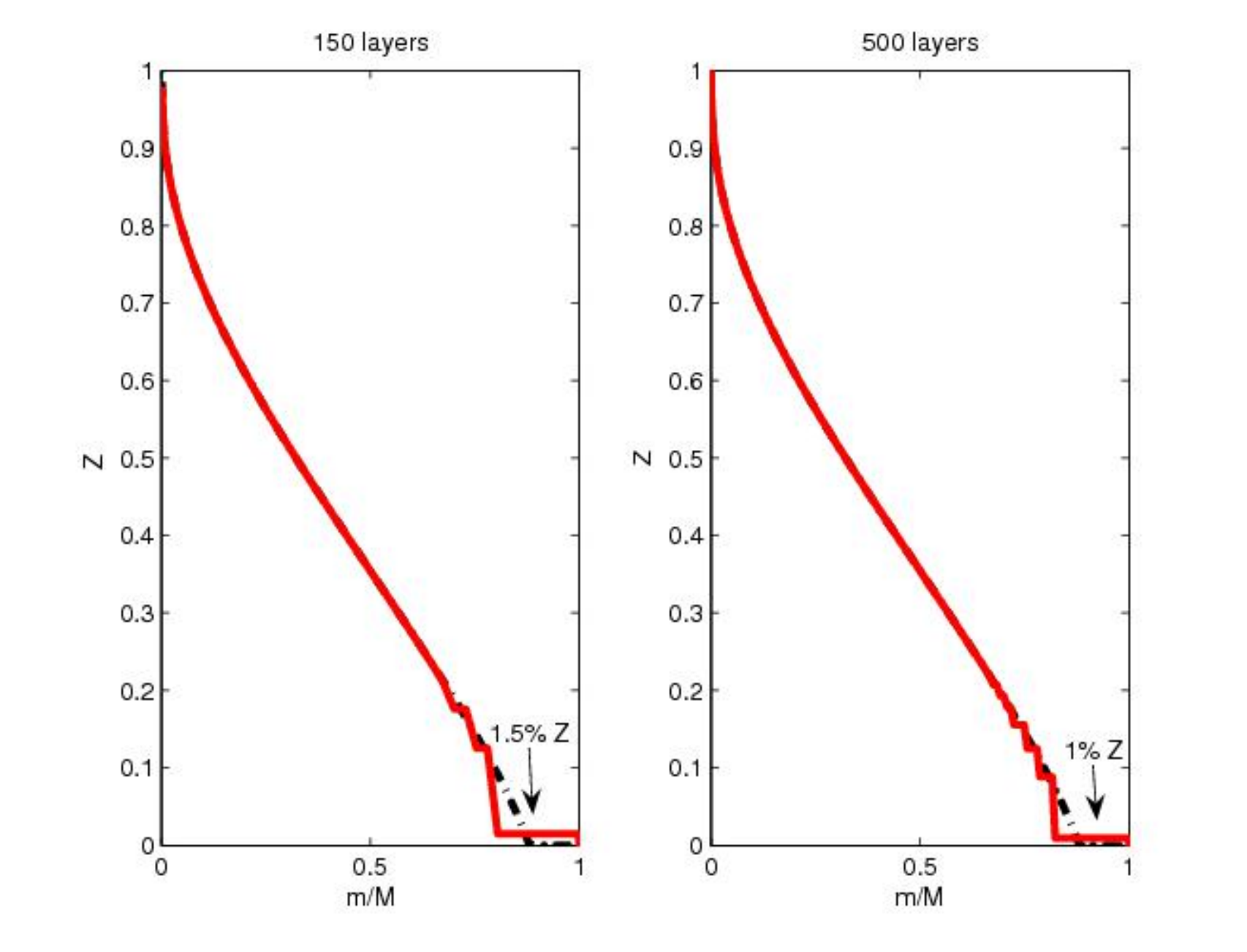}}
\centerline{\includegraphics[angle=0, width=10cm]{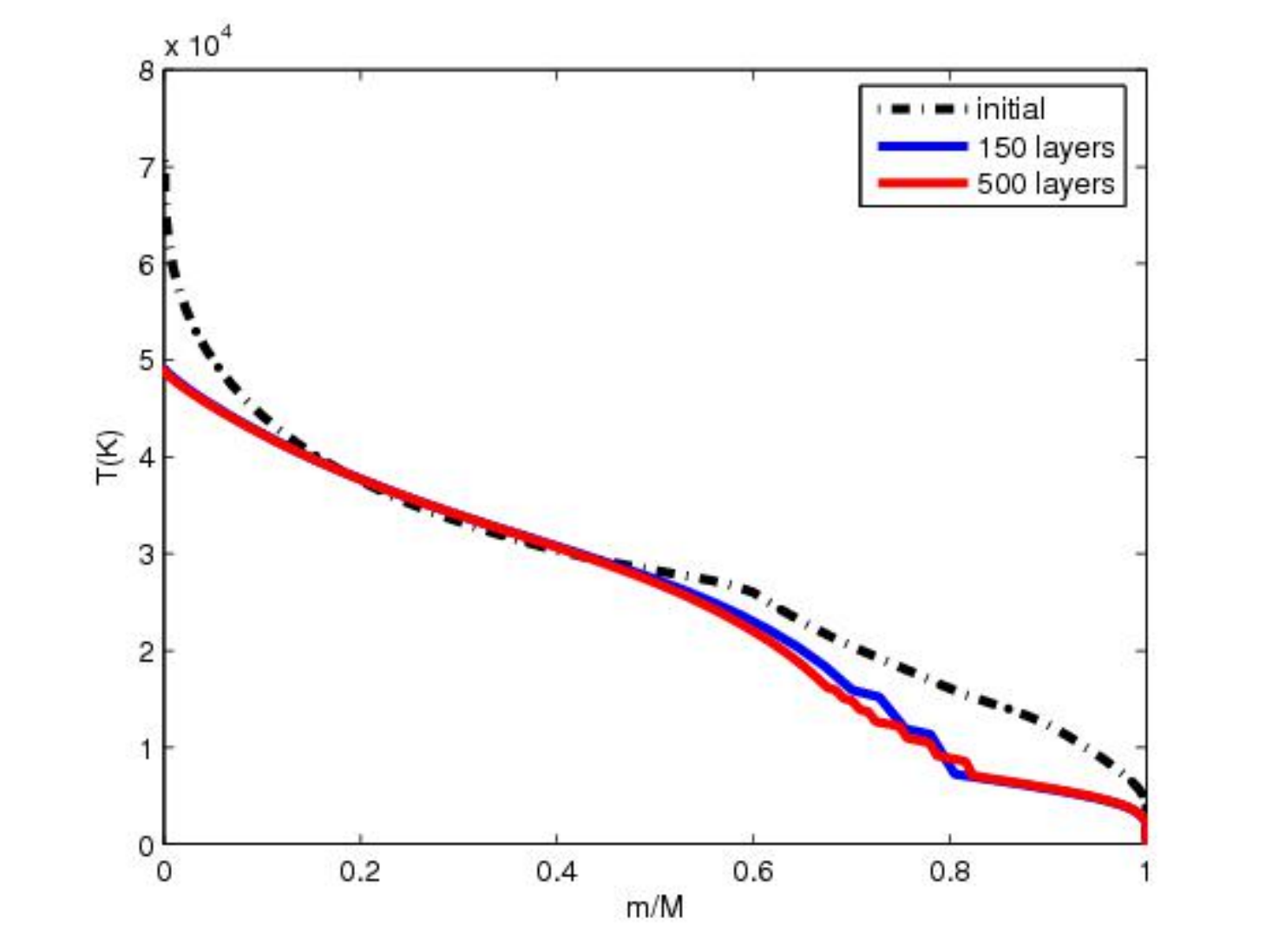}}
\caption{ Upper panel: Z as a function of normalized mass for a 1\mj planet 150 mass mesh points (left) and 500 mesh points (right). The initial structure is the same for both cases and is given by the dashed-dotted black curve. The red curves show Z after $10^{10}$ years for the two cases. Lower panel: temperature profile after $10^{10}$ yr of evolution, for the above planets with 150 (blue) and 500  mesh points. The initial temperature profile is presented also (dashed-dotted).}\label{z_layer}
\end{figure}

In order to assess the numerical stability of the results we ran the model of Fig.~\ref{z_grad} with 150 and 500 mesh points for $10^{10}$\,years.
The results are shown in Fig.~\ref{z_layer}. 
As the resolution is increased (i.e., more  mesh points) the number of steps in the high-Z material increases in the region $m/M=0.6-0.8$, although the position of these steps remains roughly the same. 
The location and extent of the radiative and convective regions are basically unchanged.  
As a result the temperature profile for both cases is essentially the same (lower panel of Fig.~\ref{z_layer}). 
The large step at the outer edge of the planet is slightly reduced. This is because the increased resolution allows us to fix the outermost convective region more precisely.  
The actual difference is relatively small, however, and the value of Z in this outer convective region decreases from 0.015 to 0.010.
As a result, the radius as a function of time remains essentially unchanged as well.  We can therefore conclude that the sensitivity of the calculated evolution to the number of mesh points is relatively small. 

\subsubsection{Dependence on composition}
\begin{figure}[ht]
\centerline{\includegraphics[angle=0, width=14cm]{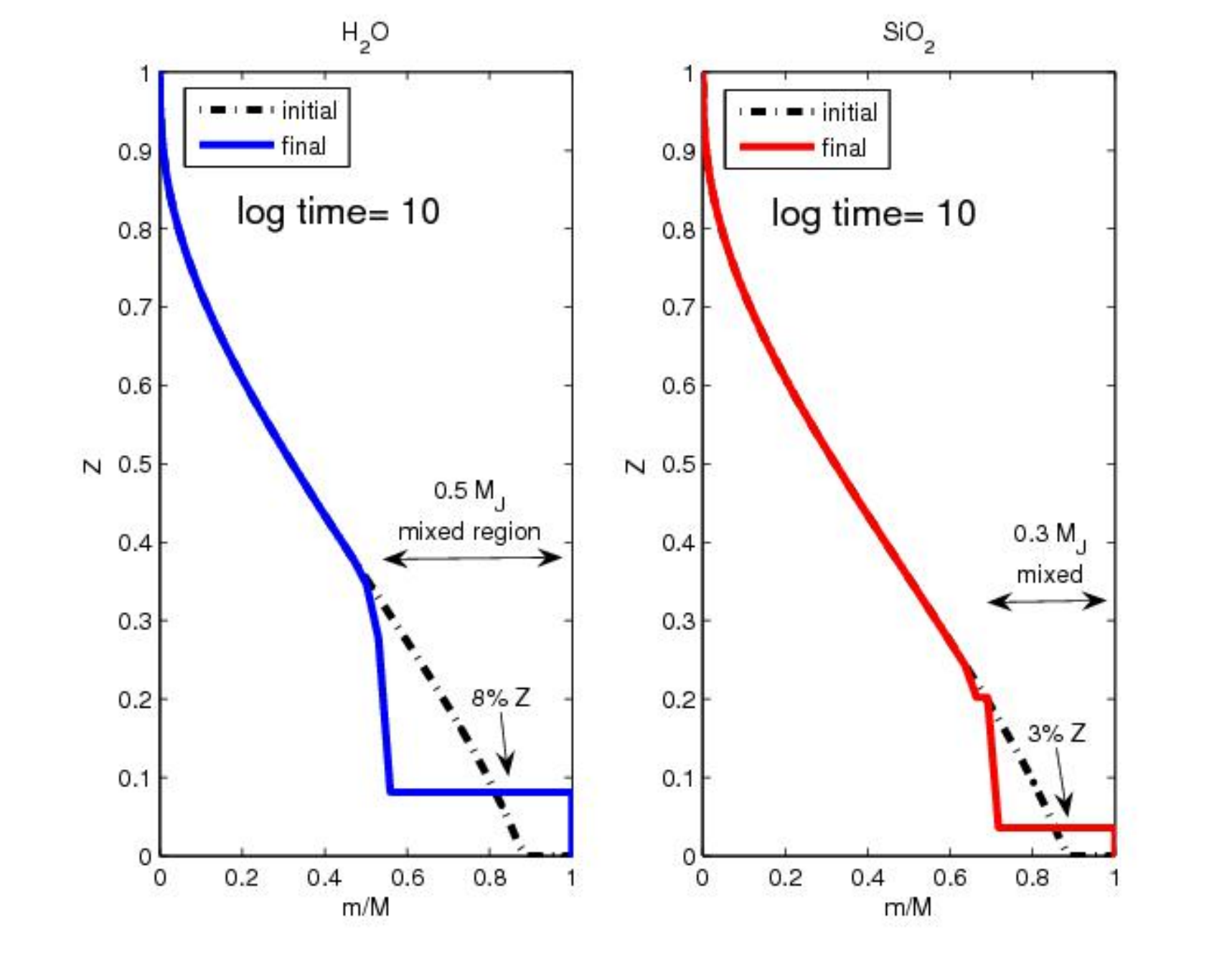}}
\caption{Z as a function of normalized mass after 10$^{10}$ years of evolution for a 1\mj planet with a total Z=0.35. The initial distribution is given by the dash-dot curve.  The final distribution is shown for \h2o (blue) and for \sio2 (red).}\label{zsh}
\end{figure}

Heavy elements in giant planet interiors can be mainly icy, or rocky. In reality, planetary interiors probably contain both. 
Necessarily, different thermodynamic properties of each material will affect the convective mixing. In addition to the known effect that the radius varies depending on whether 
rock or ice is assumed for the heavy material \citep{vazan13}, there can be an effect on the evolution via mixing.
For this reason, we compare the evolution of a 1\mj planet with an icy to one with a rocky core.  Fig.~\ref{zsh} shows the difference in mixing when using \h2o vs. \sio2 for a similar initial model. 
Here too we use $\alpha=0.5$ and start with a more gradual gradient in heavy elements. The initial high-Z distribution is shown by the dotted-black curve, 
and the distribution after 10$^{10}$ years is shown by the blue and red curves for \h2o and \sio2, respectively. As expected, mixing for the case of \sio2 is less efficient 
than for \h2o because \sio2 has a higher molecular weight.  For the case of \h2o the outer 50\% of the planet (by mass) is mixed and Z=0.08 in this region.  
For the \sio2 case only the outer 30\% of the planet is mixed and Z=0.03.
In principle, the increased metallicity of the envelope due to mixing can also affect the planetary opacity, which could in turn affect the subsequent evolution of the planet 
and its observable parameters \citep{burrows07b, vazan13}. We hope to address this topic in future work. 

\subsection{Gaussian distribution}   
\begin{figure}[ht]
\centerline{\includegraphics[angle=0, width=14cm]{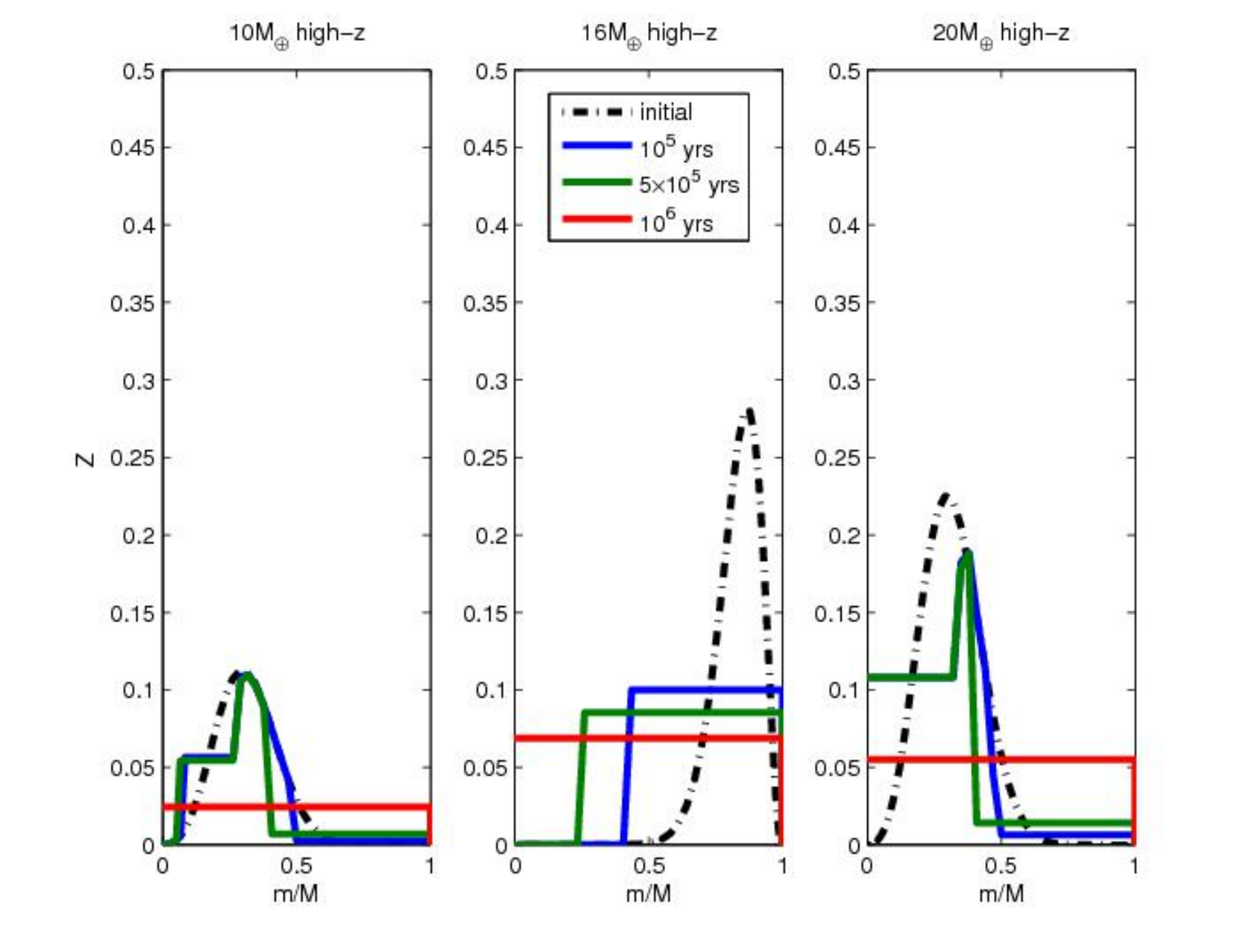}}
\caption{Z as a function of normalized mass for a 1\mj planet for different Gaussian distributions.  Left: 10\me centered at $m/M=0.3$; Center: 16\me centered at $m/M=0.8$ ; Right: 20\me centered around $m/M=0.3$.  The dashed-dotted black curves correspond to the initial distribution of the heavy elements.  Also shown are distributions after $10^5$ years (blue), $5\times10^5$ years (green) and $10^{6}$ years (red).}\label{z_gaus}
\end{figure}

The third set of cases we explore is a Gaussian distribution of high-Z material centered around some mass level in the planetary hydrogen-helium envelope during the early evolution. 
The results for this configuration for different masses and locations of high-Z material is presented in Fig.~\ref{z_gaus}. The left-hand panel presents the evolution for 10\me of \h2o injected at $m/M=0.3$.  
The dash-dot curve is the initial distribution. Mixing is found to be both downward and upward, and after $10^5$ years there is already considerable mixing (blue curve). 
After $10^6$ years, the envelope is thoroughly mixed (red curve). 

Injecting 16\me (middle panel) and 20\me (left panel) gives qualitatively similar results. In addition, we find that the location in which the material is injected does not affect 
the final results of complete mixing. 
The effect of the composition gradient on convection for these cases is small.  We compared the evolution using the \ldx criterion with that using the \swr criterion (not shown) and found very similar behavior. 
The low fraction of high-Z material (lower than Z=0.3) expected from the dissolution of even a 16\me rocky or icy object has hardly any effect on the convection criterion.  
The mixing has no real (observable) effect on the physical and thermal parameters of the planet during evolution, but it does affect the planetary envelope enrichment. 
The evolution of radius, temperature, density profiles, as well as the planetary luminosity, remains the same whether the high-Z material is originally completely mixed, or is in a Gaussian distribution around some mass point.  

\subsection{Radius and luminosity evolution}
\begin{figure}[ht]
\centerline{\includegraphics[angle=0, width=10cm]{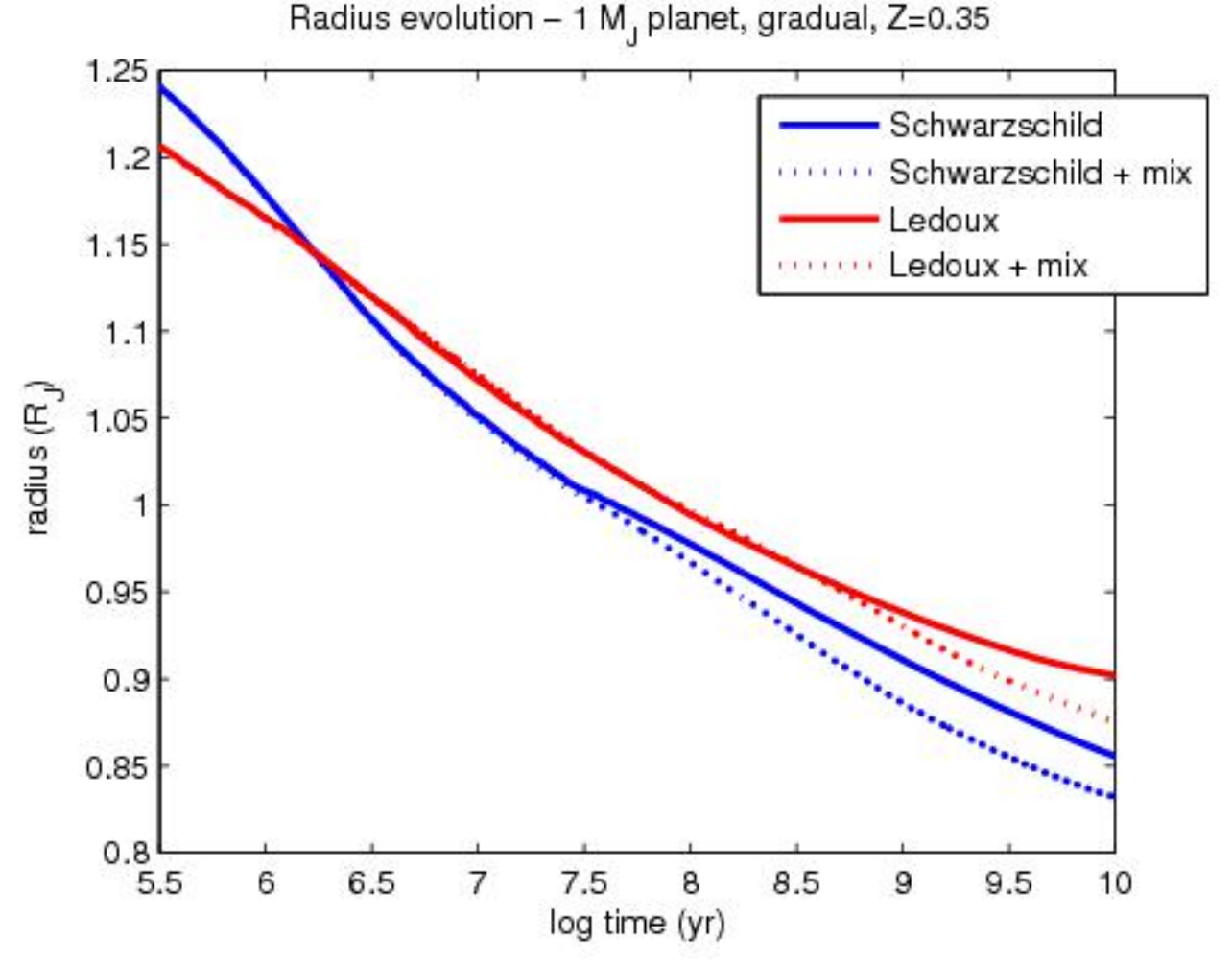}}
\centerline{\includegraphics[angle=0, width=10cm]{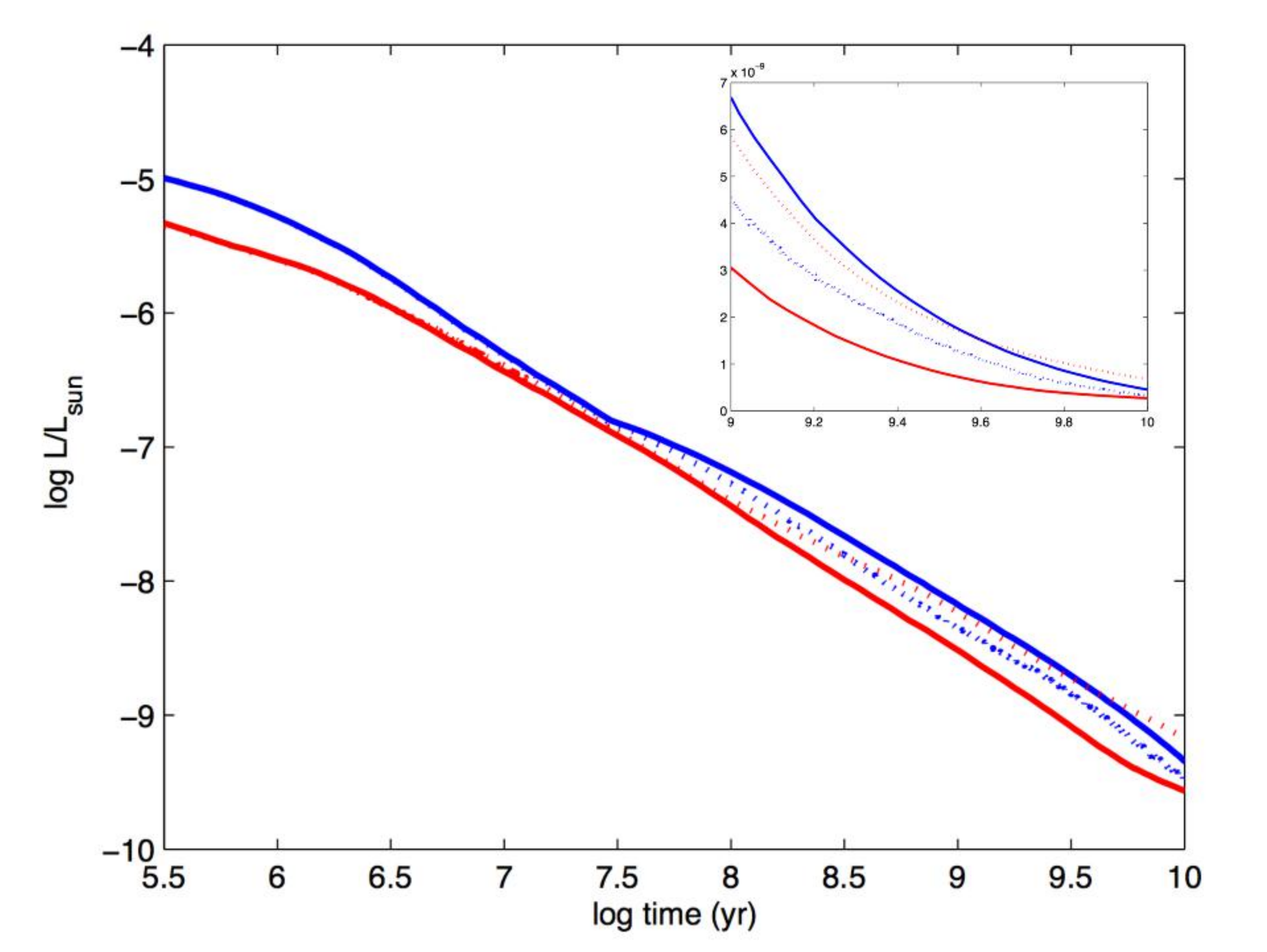}} 
\caption{Radius (top) and luminosity (bottom) as a function of time for a 1\mj planet, with Z=0.35 and an initial continuous Z-gradient (as in fig.~\ref{compgrad}, center).  Shown are evolution when using the \swr convection criterion (dotted blue) and the \ldx criterion (dotted red).  The solid curves show the evolution when there is only energy but no high-Z transport.}\label{r_grad}
\end{figure}

Since both the radius and luminosity of exoplanets can be measured, and they play key role in the characterization of the planets, we next investigate how the internal structure and mixing affect these properties. 
The change of the planetary radius (top) and luminosity (bottom) with time for a continuous compositional gradient is presented in Fig.~\ref{r_grad}. The radius is given in units of Jupiter radius (R$_J$) 
and the luminosity in units of solar luminosity, (L$_{sun}$). We compare two cases: the blue curves show the evolution when the criterion for convection is taken to be the \swr criterion 
while the red curves show the evolution using the \ldx criterion.  For each of these two cases we further consider only heat transport, that is, convection with zero diffusion coefficient $D$ (solid curves), 
and heat together with high-Z material transport (dotted curves).  
As expected, the \ldx criterion leads to slower contraction due to less efficient heat transport, and results in a larger radius. 
When material mixing is allowed ($D \neq 0$) the planetary radius decreases during the evolution, mainly due to the increased density of the envelope caused by the enrichment in heavy elements. 
The difference in luminosity between the various cases is about a factor of 2-3 throughout the evolution. 
As expected, giant planets that are modeled using the \swr criterion are found to be more luminous than planets modeled on the \ldx criterion, due to more efficient heat release. 
Since mixing occurs only at a later stage the luminosity (and radius) for the cases with and without mixing for both \swr  and \ldx is the same up to about 10$^{7.5}$ years, 
from that point on the difference in luminosity (and radius) becomes more apparent.
For both the \ldx and the \swr scenarios the mixing reduces the radius, leading to a lower luminosity. However, while the luminosity for the \swr scenario decreases when mixing is included, 
for the \ldx case it increases. This means that the effect of the slower heat release in the case of \ldx is more important than the reduction in radius (surface area) 
due to the heavier envelope.
\par

\begin{figure}[ht]
\centerline{\includegraphics[angle=0, width=10cm]{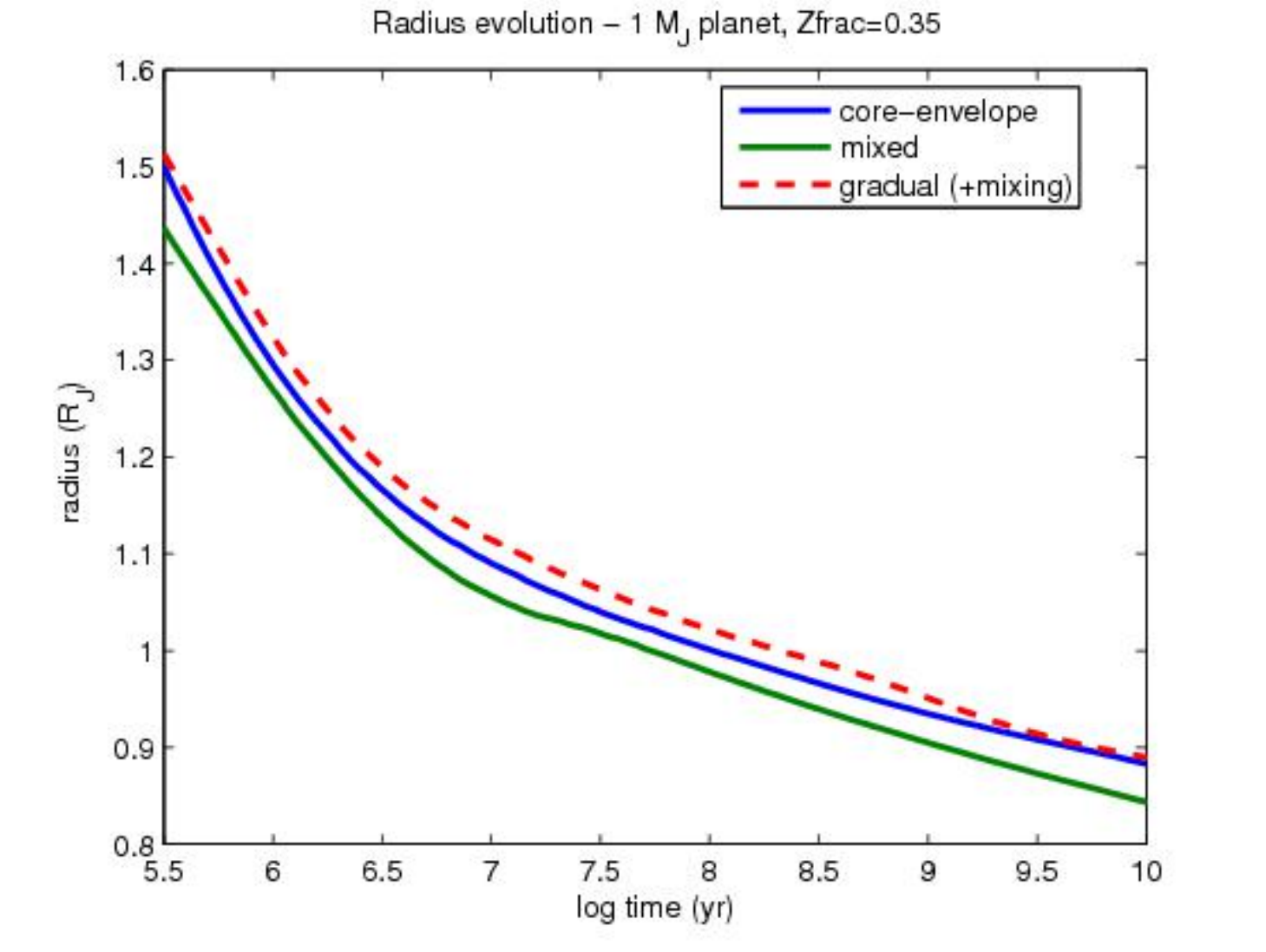}}
\centerline{\includegraphics[angle=0, width=10cm]{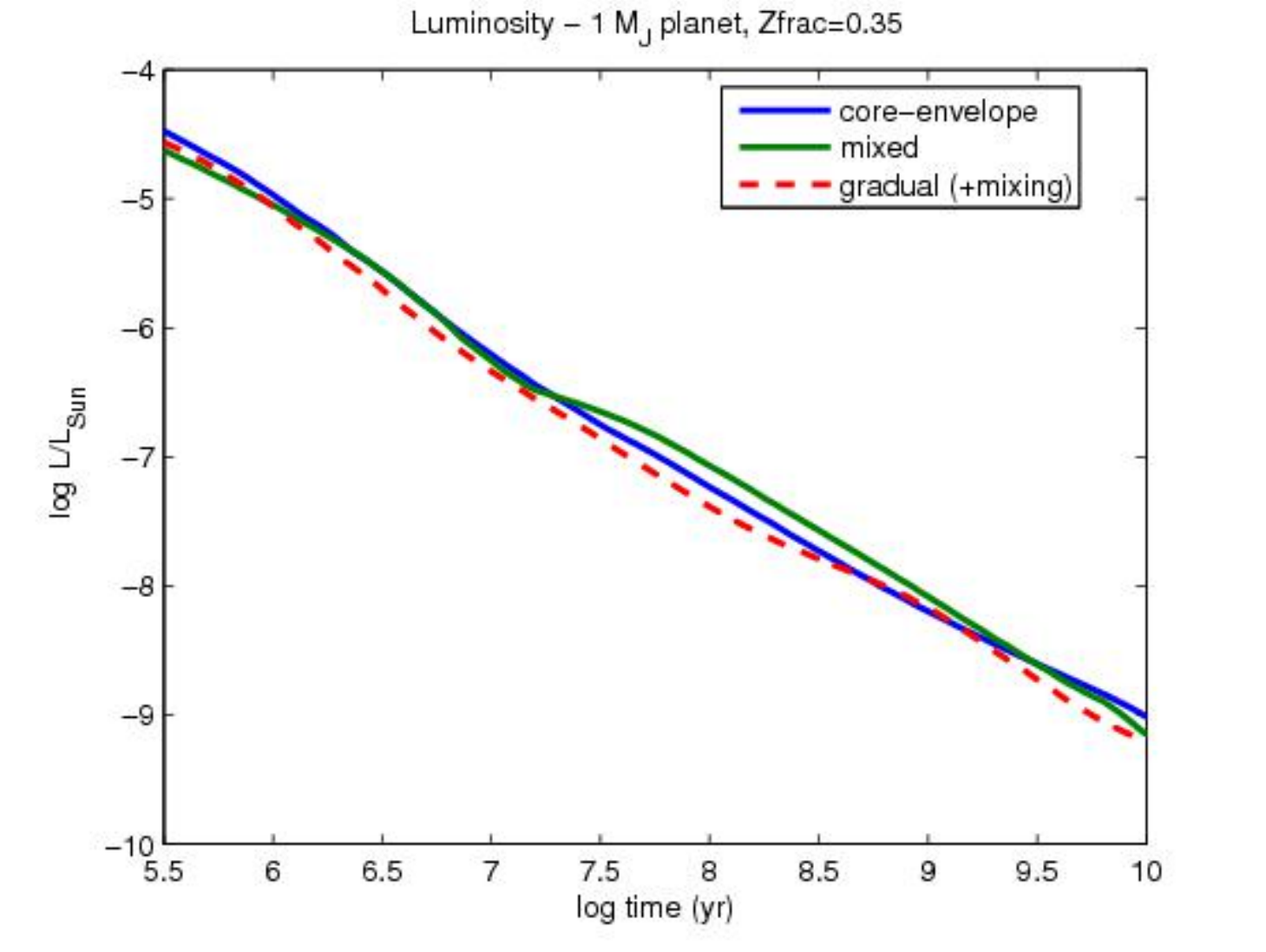}}
\caption{Radius (top) and luminosity (bottom) as a function of time for a 1\mj planet for a continuous Z-gradient with Z mixing (red dash), pure high-Z core plus solar envelope (blue), and homogeneously mixed (green). In all cases Z=0.35.}\label{r_035}
\end{figure}

The structures of core-envelope and the completely mixed planet are often considered as the extreme cases which bracket intermediate structures.  
In Fig.~\ref{r_035} the radius (top) and luminosity (bottom) evolution of planets of 1\mj, with the same initial Z fraction (Z=0.35), but different initial  distributions, is presented. 
We consider three different cases: (1) a planet with a continuous composition gradient (red-dash),
(2) an initially fully mixed planet (green), and (3) a pure high-Z core + hydrogen-helium envelope (blue).  
The fully mixed planet has large convective regions during its evolution (no compositional gradients), and heat can be rapidly transported out of the planet.  
For the core-envelope case convection is suppressed at the core envelope boundary, so that the core cools more slowly, but the envelope convects and loses heat fairly rapidly. 
Its radius stays larger than the fully mixed case, in part for that reason. However, the planet with the gradual gradient in composition has convection suppressed throughout most of the volume, 
and therefore cools more slowly than either of the previous cases.  As a result, its radius is larger that either of the two 'bracketing' cases. 
bottom panel of the figure.
The difference in planetary luminosity for the different cases, as shown in the bottom panel of the figure, are related to the heat release rate of the planets. 
The homogeneous planet, which has the most efficient heat release, has the higher luminosity during most of the evolution time. The gradual planet, which releases heat less efficiently, 
has the lower luminosity. The correlation changes after a few billion years, since the difference in radius also affects luminosity: the fully-convective homogeneous case has a smaller radius. 

The influence of mixing on the energy budget of the planet is negligible.  This is because the amount of material that mixes is relatively small, so that the amount of energy taken up by 
raising the high-Z material out of the potential well does not affect the energy of the planet. 
One might expect that the upward mixing of high-Z material would convert thermal energy into gravitational energy thus temporarily speeding up the cooling of the planet.  
Such an effect, however, if present, is small, and there is hardly any noticeable difference in the thermal energy between the mixed and unmixed case. In fact, the mixing causes 
the total radius of the planet to decrease more quickly, as can be seen in Fig.~\ref{r_grad}, and so the gravitational energy becomes more negative for the case with mixing. 

\begin{figure}[ht]
\centering
\centerline{\includegraphics[angle=0, width=10cm]{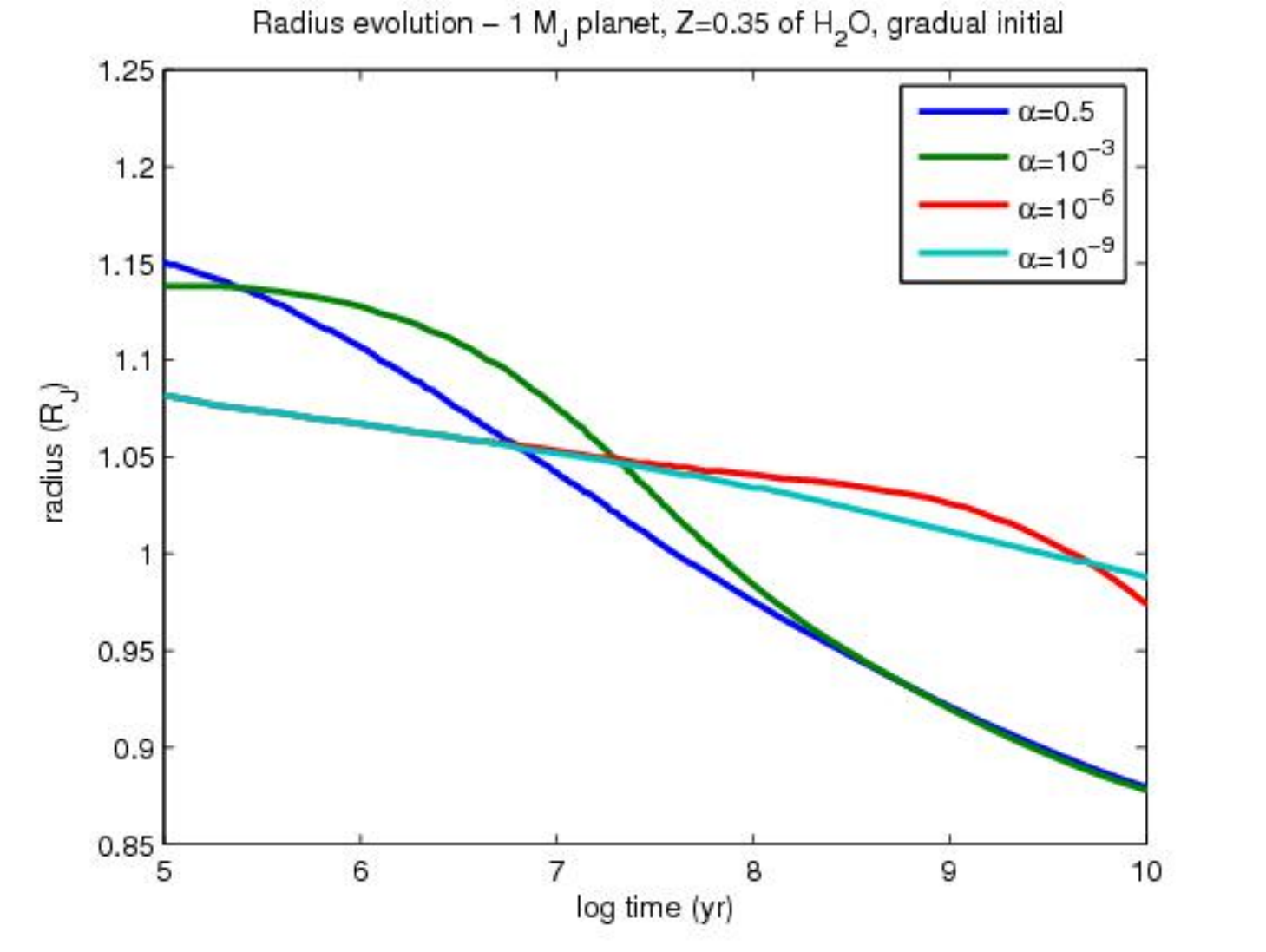}}
\centerline{\includegraphics[angle=0, width=10cm]{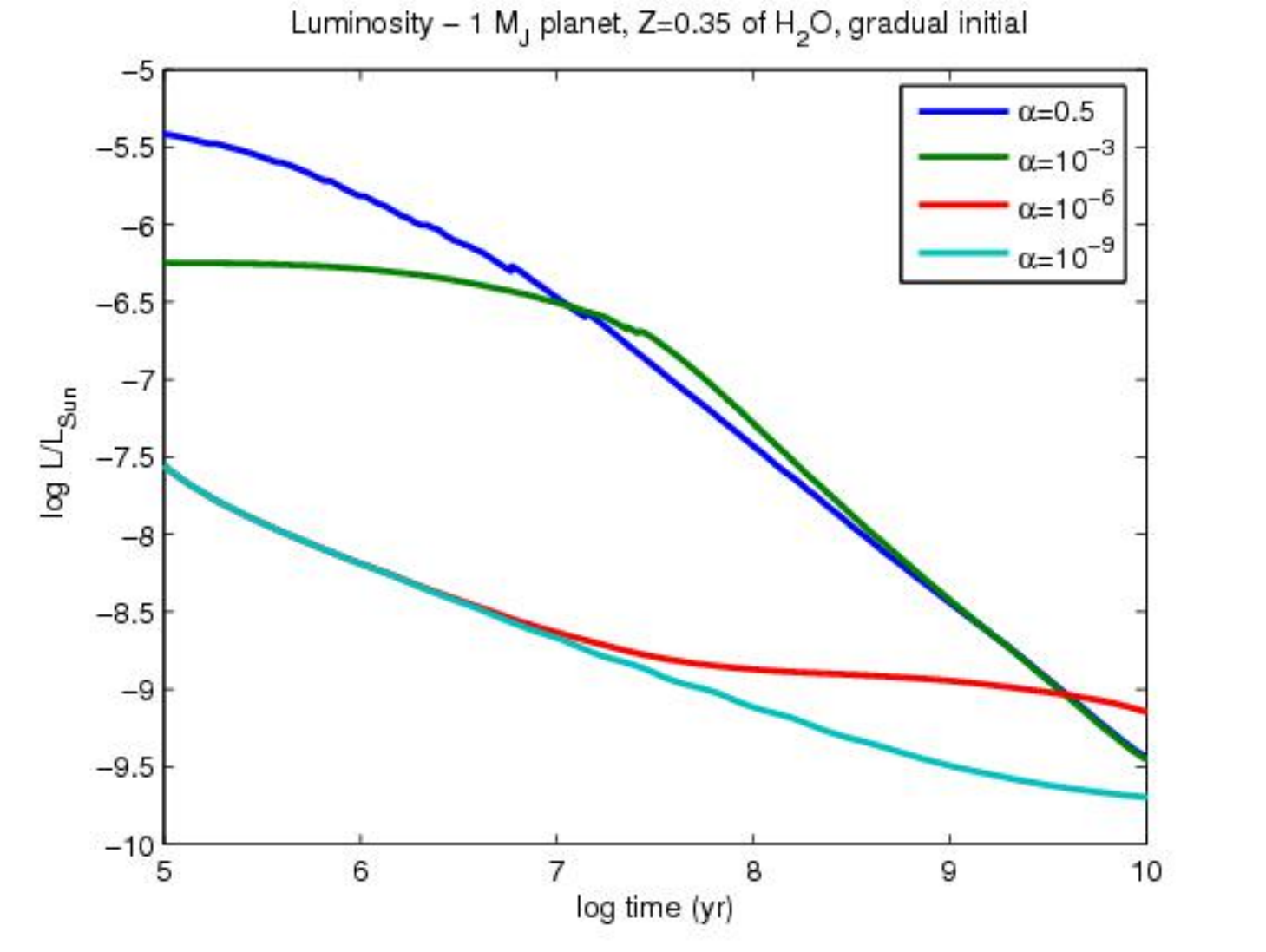}}
\caption{Radius (top) and luminosity (bottom) as a function of time for a 1\mj planet with Z=0.35 and an initial continuous Z-gradient. Different colors are for different mixing length parameters, $\alpha$. The internal high-Z distribution for those cases is presented in fig.~\ref{zalpha}.}\label{ralpha}
\end{figure}

Finally, Fig.~\ref{ralpha} shows the evolution of 1\mj planet with different mixing length parameters. Lower $\alpha$ values result in slower convection, which leads to slower cooling and 
larger radii but lower luminosities. 
Since the mixing length parameter, $\alpha$, affects the convection and mixing velocity, but not the occurrence of convection, different internal structures are developed during the evolution,
leading to different convective regions in each model. This results in different cooling histories and radius and luminosity evolution. 
Thus, there is no simple correlation between $\alpha$ and the planetary radius/luminosity, as can be seen from Fig.~\ref{ralpha}. 

It is clear from the figures that the radius and luminosity at a given time for a planet do not depend only on its composition, but are also 
dependent on the distribution of the heavy elements as well as on whether mixing of material occurs as the planet evolves. Therefore, one must be careful when using the simple mass-radius 
diagram for characterizing giant planets and inferring their bulk compositions. 
 
\section{Discussion and conclusions}
We investigate how convective transport of heat and high-Z material affects the evolution of giant planets. The evolution is followed allowing for both heat transfer and mixing of high-Z material.  
We consider various initial distributions of heavy elements, and find that the primordial internal structure has 
 an important role in determining the subsequent evolution and final internal structure. We use both the \swr and the \ldx criteria for the onset of convection. 
When the \ldx criterion is used, we find that, in the case of a distinct core-envelope boundary, core erosion is not possible due to the sharp transition in composition, which suppresses convection. 

An initial interior structure in which the heavy elements are distributed with a gradual composition gradient within the planet can become homogeneously mixed in the outer regions due to convection. 
The amount of mixed material depends on the mixing length parameter as well as on the composition of the heavy elements. 
The main effect of convective mixing is to enrich the envelope in high-Z material.  This reduces the radius of the planet during the long-term evolution. 
It is often assumed that the evolution of a planet can be bracketed by two end-member models: a planet where the high-Z material is fully mixed and a planet where the high-Z material is completely in a core.  
We have shown that a planet with a Z distribution that varies continuously within the interior can have a radius larger than both the end-members due to less efficient heat release. 
Because the details of the mixing depend on the initial composition and its depth dependence, and because these details influence the evolution and the final structure of the planet, 
we conclude that the primordial internal structure of giant planets, in particular the  initial high-Z mass distribution, is important and can affect the subsequent evolution. 

The primordial internal structure of a giant planet depends on its formation mechanism. Although both core accretion and disk instability can lead to various compositions and internal structures \citep{Helled2014}, it is often felt that giant planets formed by core accretion will have relatively massive heavy-element cores, while giant planets in the disk instability scenario are expected to be more homogeneously mixed. 
In a way, the core-envelope internal structure we consider could be a representative of core accretion while the gradual change in Z could represent disk instability planets if 
efficient planetesimal accretion takes place. However, a gradual distribution of heavy elements is possible in planets formed by core accretion, 
depending on where the heavy elements are deposited  \citep{Iaroslavitz07} and whether major mixing occurs during runaway gas accretion. 
It is therefore clear that we need to establish a better understanding of the expected composition and heavy element distribution from giant planet formation models. 
Since the primordial distribution of heavy elements depends on the formation process, and since the evolution depends on the primordial internal structure, and at the same time, 
the internal structure can change with time due to mixing and settling, a coherent picture of giant planets can be achieved only by linking the three aspects of origin, evolution, and 
internal structure self-consistently. 

\section*{Acknowledgments}
We thank P.~Bodenheimer, D.~Stevenson, G.~Schubert and N.~Nettelman for helpful remarks. 
We are also grateful to a referee for valuable comments.
This study was supported by the Israel Science Foundation grant \# 1231/10. 
AV acknowledges support from the Israeli Ministry of Science via the Ilan Ramon fellowship.

\section*{Appendix A: Ledoux's criterion and the mixing-length recipe}

Ledoux's criterion for the convective stability of a hydrostatic configuration is usually stated for 
mixtures in which the equation of state
(EOS) depends on composition through the molecular weight $\mu\,$. In an ideal gas, for example, the EOS
is $p=(R_G/\mu)\rho T\,$, where $R_G$ is the universal gas constant. In other cases, for example when 
the EOS for a mixture with mass fractions $(X_1,...,X_M)$ is
$${1\over\rho}=\sum {X_j\over\rho_j(p,T)}$$
(the additive-volume law), Ledoux's criterion requires some generalization. 
We prefer to derive it {\it ab initio.}

The criterion for convective stability is
$${d\rho\over dp} \ge {\partial\rho(p,s,X)\over\partial p} ,\eqno(A1)$$
where the derivative on the right hand side is taken at constant specific entropy $s$ and constant mass ratios $X_j\,$
(collectively denoted by $X$). On the left-hand side, we have the slope of the configuration's $(p,\rho)$ structure line. Now
$$d\rho={\partial\rho(p,T,X)\over\partial p}dp
          +{\partial\rho(p,T,X)\over\partial T}d T
          +{\partial\rho(p,T,X)\over\partial X_j}dX_j  ,\eqno(A2)$$
where the repeated index in the last term implies summation over all species\footnote{In differentiating with respect to the $X_j$'s, we disregard the fact that $\sum X_j=1$. This is guaranteed by eqs.~(5)--(6): since the diffusion coefficient $D$ is common to all species, we have $\sum F_j=0$ whenever $\sum X_j=1$ throughout the planet.
According to eq.~(5), then, $\partial\sum X_j/\partial t=0$ everywhere, and $\sum X_j=1$ will continue to be satisfied throughout. The requirement that $\sum X_j=1$ everywhere is in fact an initial condition.}. Thus, the left-hand side of (A1) is
$${d\rho\over dp}={\partial\rho(p,T,X)\over\partial p}
                        +{\partial\rho(p,T,X)\over\partial T}{d T\over dp}
                        +{\partial\rho(p,T,X)\over\partial X_j}{dX_j\over dp}.  $$
According to (A2), 
$${\partial T(\rho,p,X)\over\partial X_j}=-{\partial\rho(p,T,X)/\partial X_j\over\partial\rho(p,T,X)/\partial T} .  $$
If we replace $\rho(p,s,X)$ by $\rho(p,T,X)$, where $T=T(p,s,X)$, the right-hand side of (A1) becomes
$${\partial\rho(p,s,X)\over\partial p}={\partial\rho(p,T,X)\over\partial p}
                                      +{\partial\rho(p,T,X)\over\partial T}{\partial T(p,s,X)\over\partial p} .$$
With the standard notations
$$\nabla\equiv{d\ln T\over d\ln p} , \qquad \nabla_A\equiv{\partial\ln T(p,s,X)\over\partial\ln p} ,$$
as well as the additional notation
$$\nabla_X\equiv{\partial\ln T(\rho,p,X)\over\partial X_j}{dX_j\over d\ln p}, \eqno(A3)$$
the criterion (A1) for convective stability becomes
$${\partial\rho(p,T,X)\over\partial\ln T}(\nabla-\nabla_A-\nabla_X) \ge 0 .\eqno(A4)$$
For materials satisfying $Q\equiv-\partial\ln\rho(p,T,X)/\partial\ln T > 0$ (positive volume {\it expansion} - rather than {\it contraction} - coefficient) this becomes
$$\nabla-\nabla_A-\nabla_X \le 0, \eqno(A5)$$
which is the desired, general form of Ledoux's criterion for convective stability. In a layer of uniform composition
$\nabla_X=0$, and (A5) reduces to Schwarzschild's criterion $\nabla-\nabla_A \le 0 .$

If, as in an ideal gas, the EOS depends on the mass fractions only through the molecular weight, then
$${\partial\ln T(\rho,p,X)\over\partial X_j}{dX_j\over d\ln p}={\partial\ln T(\rho,p,\mu)\over\partial\mu}{\partial\mu\over\partial X_j}{dX_j\over d\ln p},$$
and the inequality (A5) assumes its familiar form
$$\nabla-\nabla_A-{d\ln\mu\over d\ln p} \le 0 .\eqno(A6)$$
since, in a perfect gas, $\partial\ln T(\rho,p,\mu)/\partial\ln\mu=1$.

Linear stability analysis predicts that, when the inequality (A1) is violated, small perturbations will grow exponentially with time \citep{kanikovetz67}, but the actual amplitudes 
reached lie outside the linear analysis: they are limited by the non-linear terms that were left out by that analysis.

If the total luminosity is due to radiation (and conduction),
$$L=-4\pi r^2{c\over\kappa\rho}{dp_R\over dr}, \qquad p_R={1\over3}aT^4 .$$   
The `temperature gradient' $d\ln T/d\ln p$ that results from this equation and the hydrostatic equation (\ref{h.eq}) is
$$\nabla_R={d\ln T\over d\ln p}={p\over4p_R}{\kappa L\over4\pi cGm} .\eqno(A7) $$
Now, if this temperature gradient violates the stability criterion, that is, if
$$\nabla_R-\nabla_A-\nabla_X > 0, $$  
then convection will set in: a part of the luminosity will then be carried by a convective flux,
and the rest by radiation/conduction, requiring a $\nabla$ which is less than $\nabla_R$.

For any $\nabla>\nabla_A+\nabla_X$ the excess in $d\rho/d\ln p$ of a convective element, relative to its suroundings,
is given by the left-hand side of (A4).
The density excess after travelling a `mixing-length' distance $\ell$ is obtained by multiplying by $\ell d\ln p/dr$.
The resulting density excess is therefore
$$\delta\rho\simeq -{\partial\rho(p,T,X)\over\partial\ln T}{\ell\over H_p}(\nabla-\nabla_A-\nabla_X) ,$$   
where $H_p$ is the pressure scale height, defined by $H_p^{-1}=-d\ln p/dr$.
The corresponding temperature excess then follows from eq.~(A2):
$$\delta\ln T\simeq{\ell\over H_p}(\nabla-\nabla_A) .  $$  
Actually, however, convective elements do not move adiabatically, since they continually lose energy to their surroundings by radiation or conduction. 
We therefore replace $\nabla_A$ in the foregoing formulae by the (as yet unknown) element's temperature gradient $\nabla'$:
$$\delta\rho\simeq -{\partial\rho(p,T,X)\over\partial\ln T}{\ell\over H_p}(\nabla-\nabla'-\nabla_X) ,\eqno(A8)$$
$$\delta\ln T\simeq{\ell\over H_p}(\nabla-\nabla') .\eqno(A9)$$
From ${\bf(v\cdot grad)v} \simeq {\bf g}\delta\ln\rho$ we obtain an estimate for the convective speed,
$$v^2\simeq g\ell\delta\ln\rho\simeq gH_pQ\big({\ell\over H_p}\big)^2(\nabla-\nabla'-\nabla_X) .\eqno(A10)$$
The convective energy flux is $F_c=\rho v\delta h$, where $\delta h$ is the excess heat per unit mass, given by the change in the specific enthalpy
$h=u+p/\rho$, where $u$ is the specific internal energy. 
At a given pressure, we have
\begin{eqnarray*}
dh&=&{\partial h(T,p,X)\over\partial T}dT+{\partial h(T,p,X)\over\partial X_j}dX_j \\
    &=&c_pTd\ln T+{\partial h(T,p,X)\over\partial X_j}dX_j,\end{eqnarray*}
where, again, the repeated index implies summation over all species.
Thus, according to eqs.~(A3) and (A9), the excess specific heat is
\begin{eqnarray*}
\delta h&\simeq& c_pT{\ell\over H_p}(\nabla-\nabla')+{\ell\over H_p}{\partial h(T,p,X)\over\partial X_j}{dX_j\over d\ln p} \\
&\simeq& c_pT{\ell\over H_p}(\nabla-\nabla'-\nabla_X)+{\ell\over H_p}\Big(c_pT{\partial\ln T(\rho,p,X)\over\partial X_j}+{\partial h(T,p,X)\over\partial X_j}\Big){dX_j\over d\ln p} .\end{eqnarray*}
In the last sum over compositions we expect some cancellation: after all, since $\sum X_j=1$, the $dX_j/d\ln p\,$'s cannot all have the same sign. Moreover, in a mixture of perfect gases
the sum is identically zero because each one of the coefficients of the $dX_j/d\ln p\,$'s vanishes! In what follows we shall neglect this sum and estimate the excess heat per unit mass by
$$\delta h\simeq{\ell\over H_p} c_pT(\nabla-\nabla'-\nabla_X) .\eqno(A11) $$
The resulting convective energy flux $F_c=\rho v\delta h$ is then (cf. eq.~(A10))
$$F_c\simeq (gH_pQ)^{1/2}\rho c_pT\big({\ell\over H_p}\big)^2(\nabla-\nabla'-\nabla_X)^{3/2} .\eqno(A12) $$
Two equations are needed in order to determine the unknowns $\nabla'$ and $\nabla$.
We first consider the ratio of excess energy of an element after travelling a distance $\ell$ relative to its surroundings,
which---according to eq.~(A11)--- is proportional to $(\nabla-\nabla'-\nabla_X)$, to the energy lost by radiation/conduction during the time of travel
$\ell/v$, which is proportional to $(\nabla-\nabla_A-\nabla_X)-(\nabla-\nabla'-\nabla_X)$.
An estimate \citep{mihalas78} of this ratio yields the first equation,
$${\nabla'-\nabla_A\over(\nabla-\nabla'-\nabla_X)^{1/2}}\simeq{16\sigma T^4\over\rho c_pT(gH_pQ)^{1/2}(\ell/H_p)^2\tau_H}, \eqno(A13)$$
where $\sigma$ is the Stefan-Boltzmann constant and $\tau_H\equiv\kappa\rho H_p$. The second equation is
$$F_c+F_R=F, \eqno(A14)$$
where $F_R=-c/(\kappa\rho)dp_R/dr=16\sigma T^4\nabla/(3\tau_H)$ is the radiative flux, and $F=16\sigma T^4\nabla_R/(3\tau_H)$ is the total flux.
A mixing-length recipe consists in replacing each $\simeq$ in the foregoing relations by an equality, followed by
a numerical factor. 
According to Mihalas (for example)
\begin{eqnarray*}
v&=&(gH_pQ/8)^{1/2}(\ell/H_p)(\nabla-\nabla'-\nabla_X)^{1/2} ,\\
F_c&=&(gH_pQ/32)^{1/2}\rho c_pT(\ell/H_p)^2(\nabla-\nabla'-\nabla_X)^{3/2} , \\
{\nabla'-\nabla_A\over(\nabla-\nabla'-\nabla_X)^{1/2}}&=&{32\sqrt{2}\sigma T^4\over\rho c_pT(gH_pQ)^{1/2}(\ell/H_p)^2\tau_H}\equiv 2b. \end{eqnarray*}
Eq.~(A14) and the last two equations lead to the cubic equation
$$2b'x+x^2+{3\over4b'}x^3=1 , \eqno(A15)$$
where
$$b'\equiv b/(\nabla_R-\nabla_A-\nabla_X)^{1/2} , \quad x^2\equiv(\nabla-\nabla'-\nabla_X)/(\nabla_R-\nabla_A-\nabla_X) .$$
The cubic equation is the same as the one used in the Schwarzschild context, but $b'$ and $x$ are
differently defined (except when $\nabla_X=0$). 
Finally, the solution of (A15) yields the required temperature gradient to be used in eq.~(\ref{T_eq}):
$$\nabla=\nabla_A+\nabla_X+(\nabla_R-\nabla_A-\nabla_X)(2b'x+x^2). \eqno(A16)$$
The last equation, again, reduces to the one used in the context of Schwarzschild's criterion when $\nabla_X=0$.
Efficient convection ($b' << 1$) leads to $\nabla\simeq\nabla_A+\nabla_X$; when convection is inefficient, $\nabla\simeq\nabla_R$.
The ratio of convective to total energy flux is
$${\nabla_R-\nabla\over\nabla_R}={\nabla_R-\nabla_A-\nabla_X\over\nabla_R}{3\over4b'}x^3 .\eqno(A17)$$
Finally, mixing in a convective layer is effected by eq.~(\ref{X_eq}),
where the `convective diffusion coefficient' $D$---common to all species---is a fraction of $v\ell$ (as if the convective elements were gas particles, 
moving at a mean speed $v$ with mean free path $\ell$). In this study we set $D=0.1v\ell=0.1(\ell/H_p)vH_p$.
In a radiative layer $\nabla=\nabla_R$ and $D=0$.
\par

\clearpage
\bibliographystyle{apj} 
\bibliography{allona}   
\newpage

\end{document}